\documentclass[12pt]{iopart}

\expandafter\let\csname equation*\endcsname\relax
\expandafter\let\csname endequation*\endcsname\relax
\usepackage{amsmath}

\usepackage{bm}
\usepackage{amssymb,amsfonts,latexsym,fancyhdr,graphicx,epstopdf,times,txfonts}
\usepackage{iopams}  
\usepackage{graphicx}

\usepackage{overpic}
\usepackage[english]{babel}
\usepackage{graphicx}
\usepackage {subfigure}
\usepackage[hidelinks]{hyperref}

\usepackage{xcolor}

\usepackage{verbatim}
\usepackage{lipsum}

\def\bra#1{\mathinner{\langle{#1}|}}
\def\ket#1{\mathinner{|{#1}\rangle}}

\begin{document}

\title{Statistics of orthogonality catastrophe events in localised disordered lattices}

\author{F Cosco$^1$, M Borrelli$^1$, E-M Laine$^1$, S Pascazio$^{2,3,4}$, A Scardicchio$^{5,6}$ and S Maniscalco$^{1,7}$}
\address{$^1$ QTF Centre of Excellence, Turku Centre for Quantum Physics,  Department of Physics and Astronomy, University of Turku, FI-20014 Turun yliopisto, Finland}
\address{$^2$ Dipartimento di Fisica and MECENAS, Universit\`{a} di Bari, I-70126 Bari, Italy}
\address{$^3$ Istituto Nazionale di Ottica (INO-CNR), I-50125 Firenze, Italy}
\address{$^4$ INFN, Sezione di Bari, I-70126 Bari, Italy}
\address{$^5$ The Abdus Salam International Center for Theoretical Physics, Strada  Costiera  11,  34151 Trieste, Italy}
\address{$^6$ INFN Sezione di Trieste, Via Valerio 2, 34127 Trieste, Italy}
\address{$^7$ QTF Centre of Excellence, Department of Applied Physics, Aalto University, FI-00076 Aalto, Finland}

\begin{abstract}
We address the phenomenon of statistical orthogonality catastrophe in insulating disordered systems. More in detail, we analyse the 
response of a system of non-interacting fermions to a local perturbation induced by an impurity. By inspecting the overlap between the pre and post-quench many-body ground states 
we fully characterise the emergent statistics of orthogonality events as a function of both the impurity position and the coupling strength. We consider two well-known one-dimensional models, namely the Anderson and the Aubry-Andr\'e insulators, highlighting the arising differences. Particularly, in the Aubry-Andr\'e model the highly correlated nature of the quasi periodic potential produces unexpected features in how the orthogonality catastrophe occurs. We provide a quantitative explanation of such features via a simple, effective model. We further discuss the incommensurate ratio approximation and suggest a viable experimental verification in terms of charge transfer statistics and  interferometric experiments using quantum probes.
\end{abstract}

%\maketitle

\section {Introduction}
\label{intro}

Cold atoms in optical lattices are nowadays universally accepted as an outstanding
experimental platform to realise paradigmatic models in condensed matter and high energy physics. Current and future lines of research delve deep into studying the dynamics of interacting, disordered systems \cite{bloch2012quantum,blatt2012quantum,Georgescu2014,schreiber2015observation,vardhan2017}, and lattice gauge theories \cite{zohar2013cold,banerjee2013atomic}. 
The ability to tune atomic interactions as well as potential energy profiles practically at will, enables experimentalists to simulate a variety of systems, interacting versus non-interacting, one- versus higher-dimensional and clean versus disordered. Well-known examples of models implemented in optical lattices include Heisenberg and Hubbard Hamiltonians \cite{greiner2002quantum,lewenstein2007ultracold}, and systems undergoing Anderson \cite{anderson1958absence,roati2008anderson} and many-body-localisation \cite{vznidarivc2008many,pal2010MBL,bardarson2012unbounded,de2013ergodicity,huse2013phenomenology,serbyn2013local,ros2015integrals,Imbrie2016,schreiber2015observation,brenes2018many} (for a review see \cite{nandkishore2015many,imbrie2017local}). Such direct experimental observations provide accessible and quick validation to longstanding theoretical well-known results and speculations.

In this article we investigate the collective response of some one-dimensional localised systems to local adiabatic perturbations.
Recent studies have demonstrated that a local adiabatic quench can induce a non-local rearrangement of the energy eigenstates, resulting in a non-local transfer of charge across the lattice. This effect has been confirmed in the Anderson insulator (AI) \cite{anderson1967infrared} as well as in
the Aubry-Andr\'{e} (AA) model \cite{aubry1980}, in both the interacting and non-interacting case. The effect was \emph{dubbed statistical orthogonality catastrophe} (STOC) and in Ref.~\cite{khemani2015nonlocal, deng2015exponential} it was theoretically predicted and numerically confirmed  that the typical wave function overlap $F_\textrm{typ}$ between the unperturbed ground state $|\Psi_0(0) \rangle$ and the perturbed one $|\Psi_0(\epsilon) \rangle$ ($F=| \langle \Psi_0(0) |\Psi_0(\epsilon) \rangle |$ the overlap of two many-body eigenstates for a single disorder realisation) decays exponentially
\begin{equation}
F_\textrm{typ}  \equiv \exp(\overline {\log F}) \sim  \exp (-\alpha L),
\label{stoc}
\end{equation}
in which the bar denotes the average over disorder,
$L$ is the size of the system and $\alpha$ a constant (typically of the order of $10^{-2}$ or less) that depends, among other quantities, on $\epsilon$.
This scaling behaviour is radically different from the more familiar Anderson orthogonality catastrophe, characterising metallic systems perturbed by impurities, for which the exact same overlap follows a power law decay in the system size \cite {anderson1967infrared}
\begin{equation}
F \equiv |\langle \Psi_0(0) |\Psi_0(\epsilon) \rangle |\sim L^{-\gamma} .
\label{stoc2}
\end{equation}
The difference is due to the nature of the eigenfunctions, in the first case localized, in the second extended. In absence of localization, spectral and time-dependent probes of the OC have been thoroughly discussed, and measured in experiments \cite{Goold2011,PhysRevX.2.041020,cetina2016ultrafast,schmidt2018universal}. 

For localized dynamics, the analysis in previous works was usually focused on the study of the scaling of the typical overlap in the thermodynamical (large $L$) limit. This was very reasonable for the original Anderson orthogonality catastrophe setup, in which, since the eigenstates are extended, different regions of the spectrum are statistically homogeneous and satisfy the eigenstate thermalization hypothesis \cite{Deutsch1991,Srednicki1994,RigolOlshanii}. It is also reasonable for the AI where, in one dimension, there is no mobility edge and all the states are localized \cite{anderson1958absence}, but in the AA model at large disorder, large fluctuations of $F$, when the quench site is shuffled around, yield non-trivial effects for $L$ up to a few hundreds, effectively probing different regions of the spectrum. This range of $L$ is very relevant for cold atoms experiments described in the introduction and must be taken into account if a correct description thereof is to be achieved.

Motivated by this fact, we have performed numerical studies on the non-interacting AA model, investigating both the non-local charge transfer and the statistics of orthogonality events in presence of a strong quench in the localised phase. Our main finding is the emergence of a surprising and atypical behaviour in the statistics of the catastrophe events, manifesting as a series of plateaux.  This effect will be shown to be ultimately connected to the fractal nature of the AA spectrum.

The manuscript is organised as follows: in Sec.\ \ref{model} we first briefly summarise the AI and AA models and introduce the quenching protocol and some key quantities. The results are presented in Sec.\ \ref{results}, where we also link the statistics of orthogonality events to space-energy correlations typical of a quasi-periodic potential, a key feature that is completely absent in the AI. We further discuss our results in Sec.\ \ref{discussion} and propose a possible experimental verification, before moving to conclusions in Sec.\ \ref{conclusions}.

\section{The model}
\label{model}

\begin{figure*}[!t]
 \begin{center}
\vspace {5 mm}
 \begin{overpic}{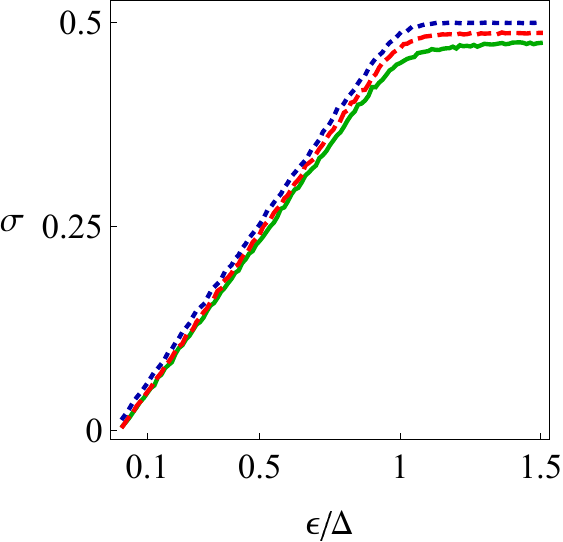} %[width = 1 \columnwidth, unit=1pt]
\put(55,100){$(a)$}
\end{overpic}
 \begin{overpic}{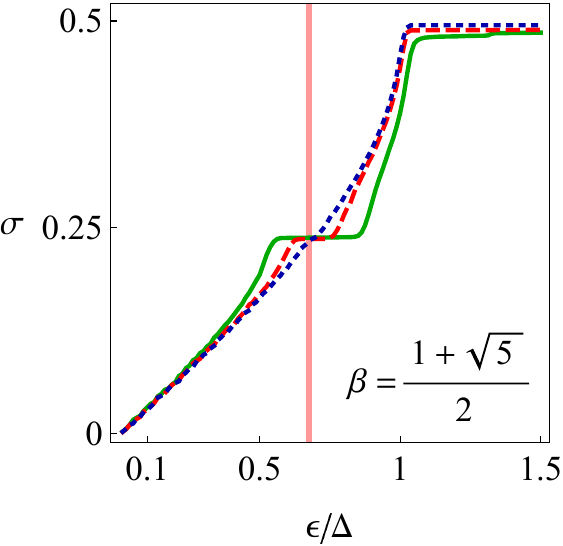} %[width = 1 \columnwidth, unit=1pt]
\put(55,100){$(b)$}
\end{overpic}

\vspace {6 mm}
 \begin{overpic}{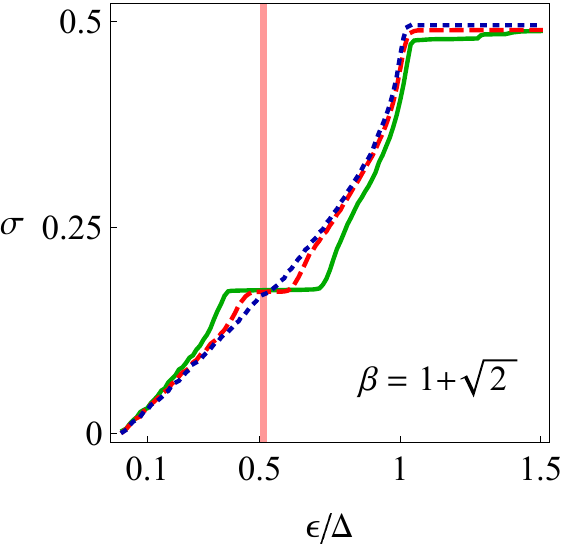} %[width = 1 \columnwidth, unit=1pt]
\put(55,100){$(c)$}
\end{overpic}
\end{center}
\caption{Probability $\sigma$ that the ground states with and without impurity become orthogonal; the impurity is located at different lattice sites and averaged over site location and 
noise (or phase of potential). $\sigma$ is displayed versus $\epsilon$ in units of $\Delta$ for $J/\Delta=0.01,0.05,0.1$ (dotted blue, dashed red and solid green respectively). The lattice size is $N_s=200$ and the two ground states are defined to be ``orthogonal" when their overlap is less than a conventional threshold $\delta=10^{-4}$. (a) AI model; (b) and (c) AA model, with $\beta$ being the golden and silver ratio, respectively. The vertical lines in (b) and (c) correspond to $\epsilon/\Delta=|\sin(2 \pi\beta)|$.}
\label{sigmaplot}
\end{figure*}

We investigate a system of non-interacting fermions in a one-dimensional  lattice subject to 
a local quench of the confining potential. The tight-binding  Hamiltonian $H$ reads 
\begin{equation}
H=-J \sum_{j=1}^{N_{s}-1}(a_{j}^{\dagger}a_{j+1}+\textrm{h.c.})+\sum_{j=1}^{N_{s}} V_{j}a_{j}^{\dagger}a_{j}+\epsilon (t) a_{x}^{\dagger}a_{x} ,
\label{TotHam}
\end{equation}
where $a_{j},a_{j}^{\dagger}$ are the $j$th site fermionic annihilation and creation operators, $J$ is the hopping parameter, $V_{j}$ the $j$-th site local potential and $x$ ranges between 1 and $N_{s}$, the length of the chain. The last term can be viewed as the effective time dependent density-density interaction with a localised impurity at position $j=x$ that is adiabatically switched on. We will study a half-filled lattice, with a total number of particles $N=N_{s}/2$, or alternatively a filling fraction $n=N/N_s=1/2$, so that the ground state will be a Fermi sea occupying half of the spectrum. At half filling the Fermi energy will be $E_F=0$.
The time evolution of the impurity potential is such that $\epsilon(0)=0$ and $\epsilon(\infty)=\epsilon$. According to the adiabatic theorem, for a system initially prepared in the many-body ground state of  Hamiltonian~\eqref {TotHam} with no impurity potential, the asymptotic final state following the adiabatic coupling will be

\begin{equation}
|\Psi_0(x, \epsilon) \rangle  = U(+\infty) | \Psi_0(\epsilon=0) \rangle,
\end{equation}
where $|\Psi_0(x, \epsilon) \rangle$ is the many-body ground with the perturbation at site $x$, as in Eq.\ (\ref{TotHam}).

Depending on the potential $V_{j}$, the Hamiltonian~\eqref {TotHam} operator can capture different models. Here, we focus on two models: a quasi-periodic potential characterised by the
profile $V_{j}=\Delta\cos(2\pi\beta j+\phi)$, with $\beta$ irrational and known in the literature as the Aubry-Andr\'e (AA) model \cite{aubry1980}, and a completely random $V_{j}$ whose local amplitudes are sampled with uniform probability in the interval $[-\Delta,\Delta]$, which yields the Anderson Insulator (AI) model \cite{khemani2015nonlocal}. The AA model is not analytically solvable, however, if $\beta$ is irrational, the resulting potential is quasi-periodic and for $\Delta =2 J$  a transition from delocalised to localised eigenstates occurs \cite{aubry1980,modugno2009,thouless1983}. On the other hand, the AI model also exhibits eigenstate localisation for any $\Delta > 0$ in 1D.

In this article we will characterise the adiabatic many-body response to a local quench by studying the overlap (or fidelity) between the pre and post-quench ground states, 
\begin{equation}
F(x, \epsilon)= |\langle \Psi_0(\epsilon=0) |\Psi_0(x,\epsilon) \rangle | .
\label{gsfidelity}
\end{equation}
We expect this quantity to vanish for any position of the impurity in the limit of very large values of $\epsilon$, when the corresponding term in Eq.~\eqref{TotHam} becomes dominant.
The overlap $F$ will be averaged over two instances. The first one is the position $x$ of the impurity potential and the second one is the noise realisation.  In the AA model, the latter corresponds to averaging over different phases $\phi$, while in the AI such an average is performed over the different random realisations of the potential $V_j$. With a slight abuse of notation, both  averages will be henceforth indicated with brackets. We therefore define 
\begin{equation}
\sigma (\epsilon) = \left\langle \frac{1}{N_s} \sum_{x=1}^{N_s} \theta(\delta - F(x,\epsilon)) \right\rangle_\textrm{noise},
\label{measure}
\end{equation}
with $\theta$ being the Theta function and $\delta$ being conventionally set at $10^{-4}$. This is the probability for an orthogonality event $F \simeq 0$ to occur and it is the key quantity to be investigated. 
We stress that this orthogonality event is due to a rearrangement of the single particle energy eigenstates, leading to an adiabatic charge transfer as shown in  Ref.~\cite{khemani2015nonlocal, deng2015exponential}.
We will always focus on the case $J \ll \Delta$, i.e.\ when the single particle eigenstates are strongly localised.  Thus, changing the impurity position within the lattice will result in an effective interaction with different levels of the single particle spectrum.

\section {Results}
\label{results}

\begin{figure*}[!t]
 \begin{center}
 \begin{overpic}[width = 0.5 \columnwidth, unit=1pt]{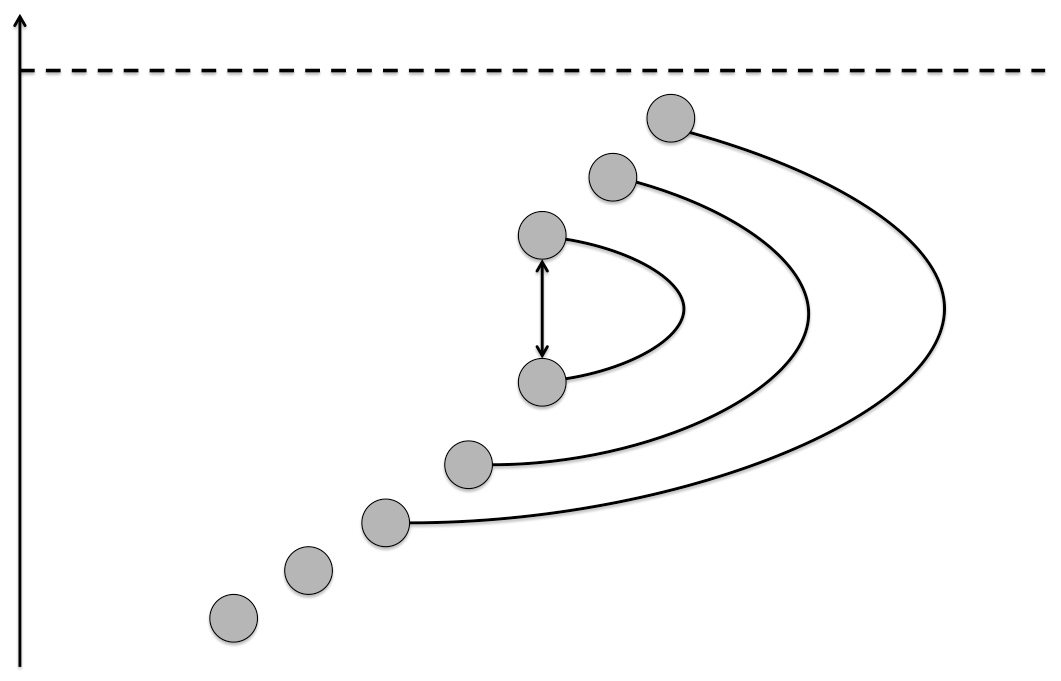}
\put(42,33){$2J$}
\put(90,60){$E_f$}
\put(-5,2){$E_n$}
\put(50,60){$(a)$}
\end{overpic}

\vspace {8 mm}
 \begin{overpic}[scale=0.95]{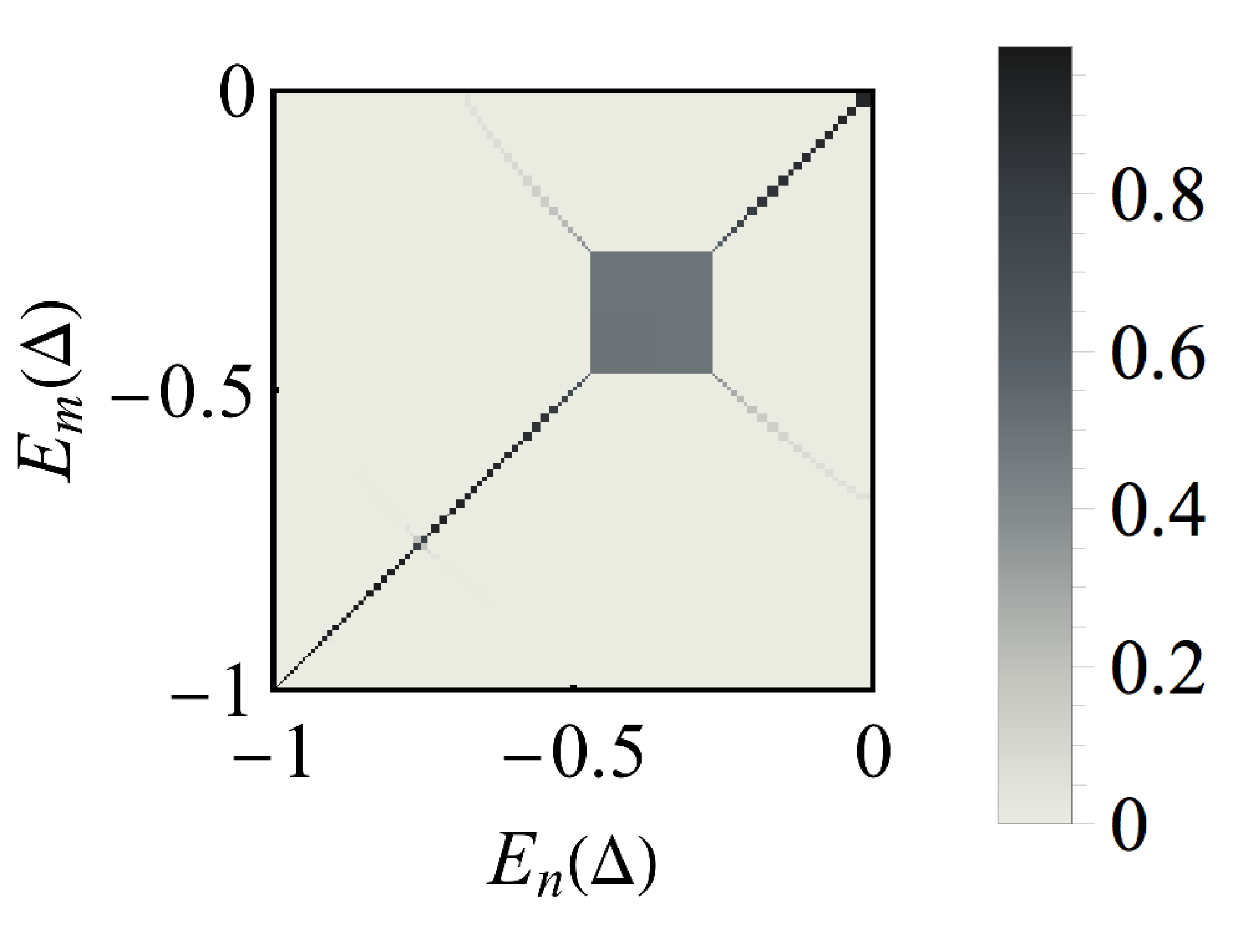}
\put(70,75){$C(E_n-E_m)$}
\put(42,75){$(b)$}
\end{overpic}
%\vspace {1 mm}
 \begin{overpic}[scale=0.95]{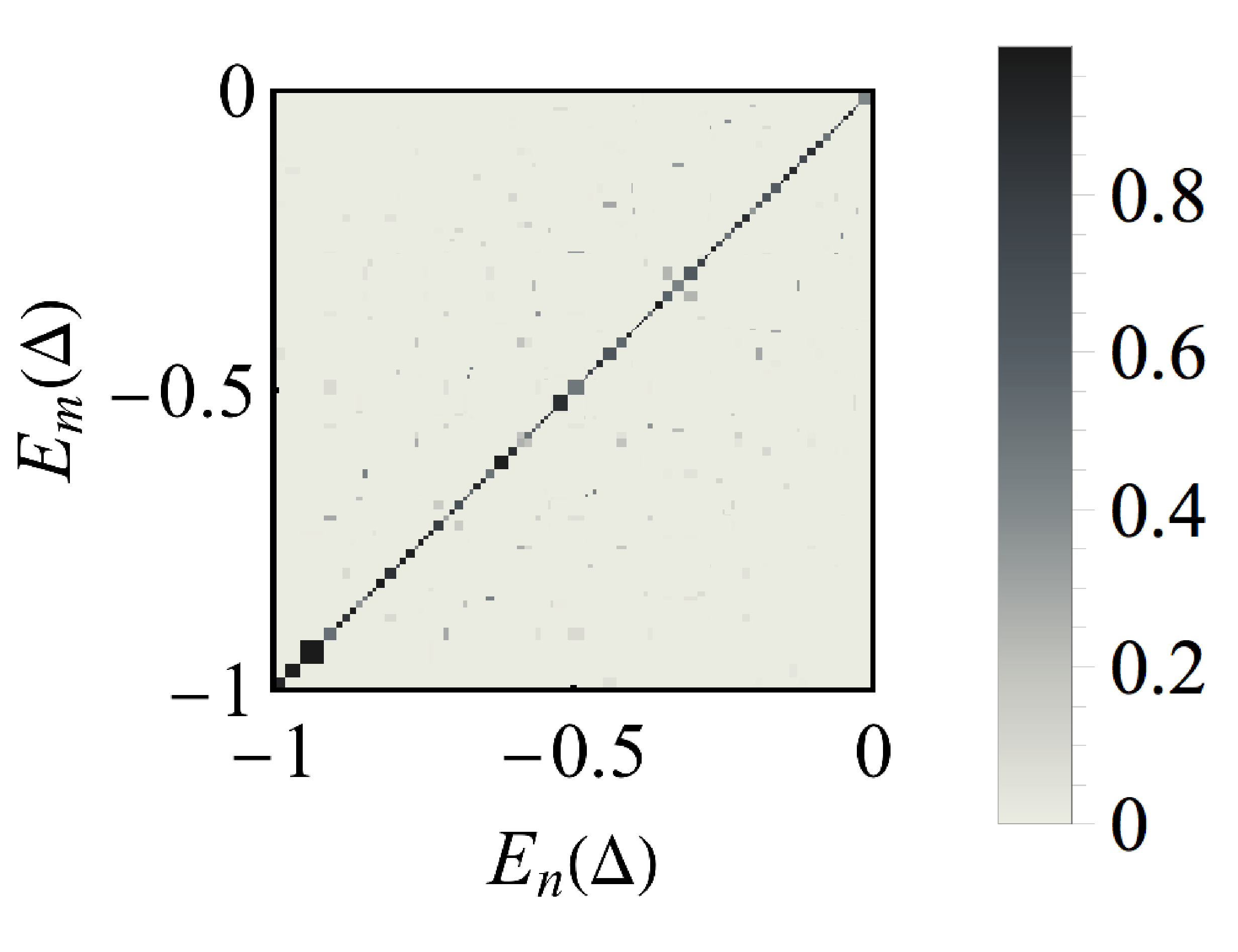}
\put(70,75){$C(E_n-E_m)$}
\put(42,75){$(c)$}
\end{overpic}
\end{center}
\caption{(a) Resonant states around the gap; the gap is of order $2 J$. Correlation function $C_{nm}$ for the (b) AA and (c) AI models. In the first case there is the clear emergence of an ordered pattern, with non-zero elements connecting states on opposite sides of the energy gaps. This feature is due to the fact that the Aubry-Andr\'e model entails a form of highly correlated noise that displays correlations in energy and space. This does not happen in the Anderson insulator, in which the non zero off-diagonal elements are randomly distributed.}
\label{correlatedstates}
\end{figure*}
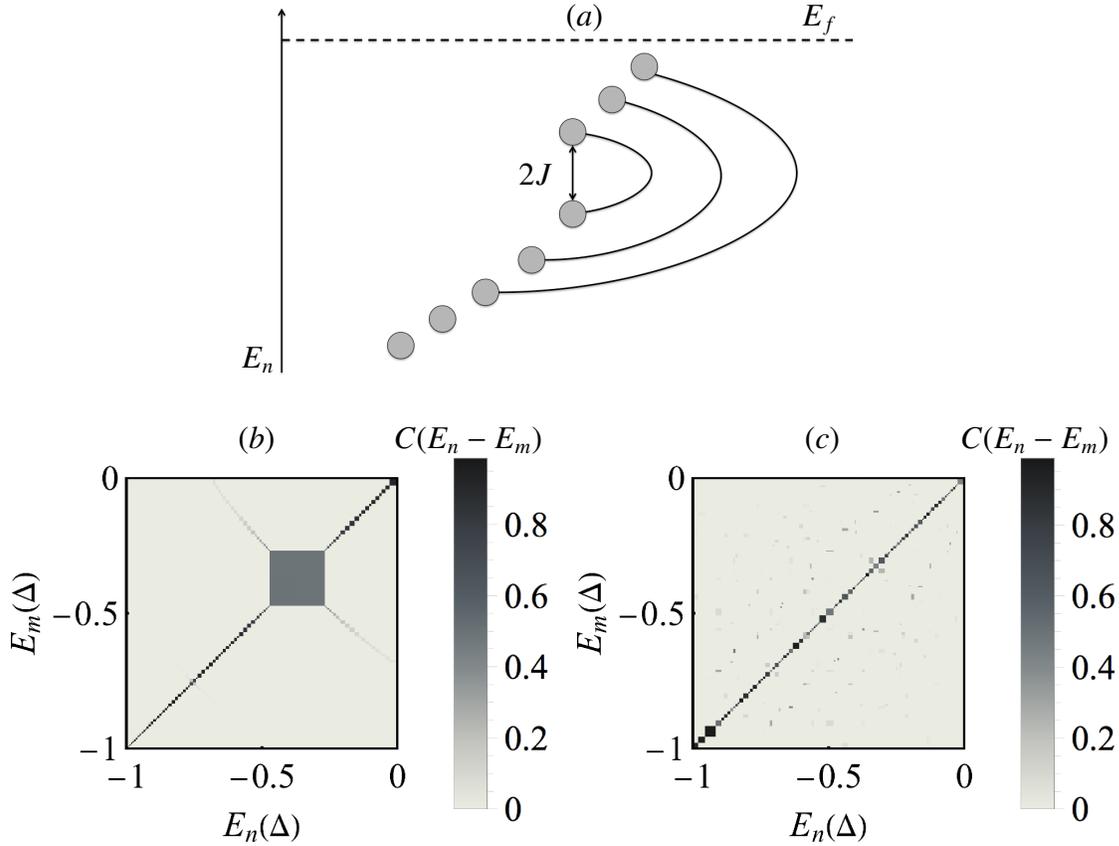

The $\sigma$ function, displayed in Fig.\ref{sigmaplot}, shows striking differences in the two models here considered. As expected, by increasing the value of the interaction between the impurity and the surrounding fermionic gas, the number of the orthogonality events increases monotonically. When $\epsilon \gg \Delta$ this saturates to $1/2$, being naturally bounded by the filling factor. This is reasonable, since in the strongly interacting regime the gas-impurity interaction strength overcomes the energy scales given by both the on-site potential and the kinetic term. Interestingly, while in the AI such saturations is achieved monotonically and without any particular structure nor dependence from the hopping parameter, in the AA model the appearance of a plateau can be clearly noticed. The amplitude of this plateau is comparable with the principal energy gap present
in the AA energy spectrum, suggesting a possible link (see Fig.\ \ref {finitesizeplot}(d)). However, since the number of events at which the plateau starts is well above the number of states between the Fermi energy and the energy gap itself, in order to understand this behaviour we need to analyse the mechanism responsible for the energy gap as well as the properties of the neighbouring eigenstates.
Since we are interested in the localised phase of the AA model, {\it i.e.} $\Delta>2J$, the spectrum is, for the most part, well approximated by the on-site potential energy 
\begin {equation}
\label{eq:onsite}
E_i \approx \Delta \cos (2 \pi \beta i+\phi ). 
\end {equation}
The only exception is when two adjacent sites have an energy difference of the same order or lower than the hopping parameter, that is $|E_{i+1}-E_i|\lesssim J$. In this case the two levels are quasi-resonant and a fermion is therefore delocalised between both the two sites $i$ and $i+1$.
This quasi-resonance condition $|E_{i+1}-E_i|\lesssim J$ leads to
\begin {equation}
\label{eq:resonant_i}
|\sin(2 \pi \beta (i+1/2) +\phi)| \lesssim \frac {J}{2 \Delta \sin (\pi \beta)}.
\end {equation}
The center of the gap is obtained by finding the states which are exactly resonating, therefore the following condition
\begin{equation}
2\pi\beta i+\phi=-\pi\beta \mod \pi
\end{equation}
yields $E_i\simeq E_{i+1}\simeq E_{g_2}$ and one finds, from Eqs.\ (\ref{eq:onsite})-(\ref{eq:resonant_i}),
that the gap is located around the energy
\begin{equation}
E_{g_2} = \pm \Delta \cos (\pi \beta).
\label{eg2}
\end{equation}
See Fig.\ \ref{correlatedstates}(a).

Two non-exactly resonant sites $i$ and $i+1$ can be described by an effective two-site Hamiltonian
\begin{equation}
H_2=
  \begin{bmatrix}
    E_i &  -J \\
    -J & E_{i}+\delta E 
  \end{bmatrix},
  \label{h2eff}
\end{equation}
where we wrote $E_{i+1}=E_i+\delta E$. To be concrete, when $i$ satisfies (\ref{eq:resonant_i}), $\delta E=0$ and $E_i=E_{g_2}$ in (\ref{eg2}).

For any state $\Psi =\sum_{j=1}^{N_s} \psi (j) \ket j$, where $\ket j = a^\dagger_j \ket {vac}$, we can use as a measure of localisation the inverse participation ratio
\begin{equation}
\mathrm {IPR}(\Psi) = \frac {1}{ \sum_{j=1}^{N_s} |\psi(j)|^4}.
\end{equation}

\begin{figure*}[!t]
\centering
 \begin{overpic}{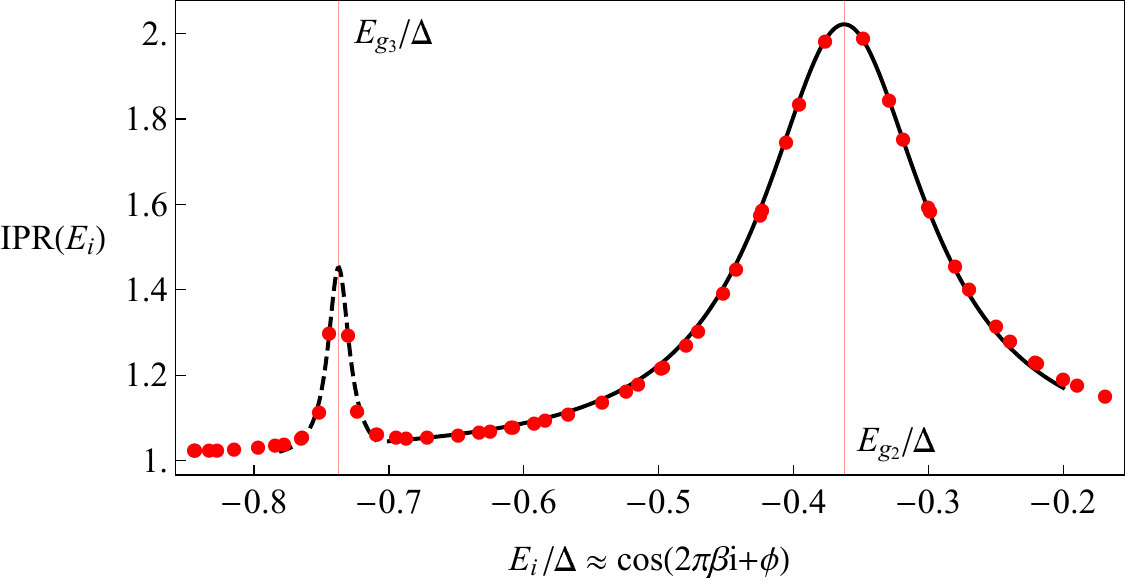}
\end{overpic}
\caption{ Inverse participation ratio defined as $\mathrm{IPR}(E_i)=1/\sum_{j=1}^{N_s} |\psi_{i}(j)|^4$, with $\psi_{i}$ eigenstate corresponding to energy $E_i$. The black curves are Lorentzian fits centered around the first two gaps $E_{g_2}$ (solid) and $E_{g_3}$ (dashed). The IPR and the eigenenergies $E_i$ are calculated for a system with $J=0.1 \Delta$, $N_s=200$ and $\beta= \frac {1+\sqrt{5}}{2}$ and a random value of the phase $\phi$. 
}
\label{lorentz}
\end{figure*}

For approximate resonances, the effective model (\ref{h2eff}) gives as IPR:
\begin{equation}
I_2 = \frac {\delta E^2+4 J^2}{\delta E^2+2J^2},
\label{ipr2}
\end{equation}
which is a Lorentzian curve with maximum 2 and width $J$ as a function of $\delta E$. See Fig.\ \ref{lorentz}.
The width of the gap is given by the energy difference between these two states, which is $2J$. See Fig.\ \ref{correlatedstates}.

One can extend this argument to states which are 2-site distant, say $i$ and $i+2$. However, in this case the required energy difference must be less than $J^2/\Delta$, as the intermediate state $i+1$ has energy of $O(\Delta)$ under which the particle has to tunnel
\begin{equation}
|E_{i+2}-E_{i}|<J^2/E_{i+1}\simeq J^2/\Delta,
\end{equation}
so
\begin{equation}
2\Delta|\sin(2\pi\beta(i+1)+\phi)\sin(2\pi\beta)|<J^2/\Delta,
\end{equation}
which leads to a gap of size $J^2/(2\Delta\sin(2\pi\beta))$ located around 
\begin{equation}
2\pi\beta i+\phi=-2\pi\beta \mod \pi
\end{equation}
and therefore the centeres of the gaps, whose size is $J^2/\Delta$ are at
\begin{equation}
E_{g_3}=\pm\Delta\cos(2\pi\beta).
\end{equation}
Notice that also these states are located around these gaps and they are delocalized on the sites $i$ and $i+2$ only. See the smaller Lorentzian in Fig.\ \ref{lorentz}.
This is the onset of the fractal structure of the gaps, with widths $O(J^{n+1}/\Delta^n)$ and located around $\Delta\cos(n\pi\beta)$. For finite $N$ however, only the first few gaps will be visible (an approximate condition is $N J^{n}/\Delta^{n}\gtrsim 2$, for at least two states have to be resonant to observe the gap of order $n+1$). 

An alternative way to visualize these resonances is through the following correlation function
\begin{equation}
C(E_m- E_n)= \sum_{j=1}^{N_s} |\psi_{n}(j)|^2|\psi_{m}(j)|^2,
\label{correlation_funciton}
\end{equation}
where $\psi_{n/m}$ are the eigenfunctions of the Hamiltonian with no impurity, i.e.
\begin {equation}
 H(\epsilon=0) \psi_n  = E_n \psi_n.
\label{eigenssytem}
\end{equation}
Figure \ref {correlatedstates}(b) shows how states on opposite sides of the main gap display a strong degree of correlation, being close in space (namely nearest neighbours).
As mentioned before, these energy gaps are the reason for the plateau structures displayed in $\sigma$, as we are now going to explain in detail. 

First of all, it must be noticed that when adding the impurity energy $\epsilon$ on the \emph{occupied} site $x$ (so $E_x<E_F=0$) the energy of the particle simply moves to
\begin{equation}
E'_x\simeq E_x+\epsilon.
\end{equation}
Whenever $E'_x>E_F=0$ an \emph{orthogonality event} occurs. However, because of the non-zero tunnelling ($J \neq 0$), the impurity energy on site $x$ affects the energies related to the other sites $E'_{x+1}, E'_{x+2},...$. As $\epsilon$ increases the condition $E'_{x+1}>E_F=0$ becomes relevant to generate new orthogonality events whenever the neighbour of the perturbed site is \emph{occupied}. The previous discussion about the distribution of the resonant states around the main gap guarantees that the states on opposite sides of the gap are nearest neighbours. This explains why the number of events at which we reach the plateau is well beyond the number of states between the Fermi energy and the energy gap itself. The last  \emph{occupied} site with an \emph{occupied} neighbour to generate an orthogonality event before the plateau is obviously paired with the site close to the Fermi energy. Being $E_F = 0$ we can assume the energy of the highest energy \emph{occupied} state to be $E \simeq 0 $ and therefore the potential on this site to be $\Delta \cos (2 \pi \beta j+\phi) \approx \Delta  \cos (m \frac {\pi}{2})$  with $m$ odd integer. 

 As a consequence the energy difference between this site and its nearest neighbour is $\delta E = \Delta | \sin (2 \pi \beta)|$, and identifies in turn the pair of states in the tails of the main Lorentzian in Fig.\ \ref{lorentz} below the Fermi energy. This $\delta E$ therefore predicts the centre of the plateau. 
In fact, the \emph{absence of orthogonality events} condition is given by
\begin{equation}
\Delta|\sin(2\pi\beta)|+J \gtrsim \epsilon \gtrsim \Delta|\sin(2\pi\beta)|-J,
\end{equation}
in our numerics $|\sin(2\pi\beta)|=0.67...$ for the case of $\beta$ being the golden ratio. In panels (b) and (c) of Fig.\ \ref{sigmaplot} we can see how this condition correctly predicts the centre of the plateau for different hopping parameters for the two cases considered, i.e. the golden ratio or the silver ratio taken as incommensurate frequencies. 
This manifests itself as a plateau in $\sigma(\epsilon)$ of width $2J$. From our previous discussion on the presence of other gaps of width $O(J^{n+1}/\Delta^n)$ arranged in a fractal structure, we can deduce that $\sigma(\epsilon)$ too will have a fractal structure, much alike a \emph{devil's staircase}.

It goes without saying that only the first few steps of the staircase are visible, because of the presence of a resolution cut due to the system size.

\section{Discussion}
\label{discussion}

\begin{figure*}[!t]
\begin{center}
\vspace {5 mm}
 \begin{overpic}[scale=0.96]{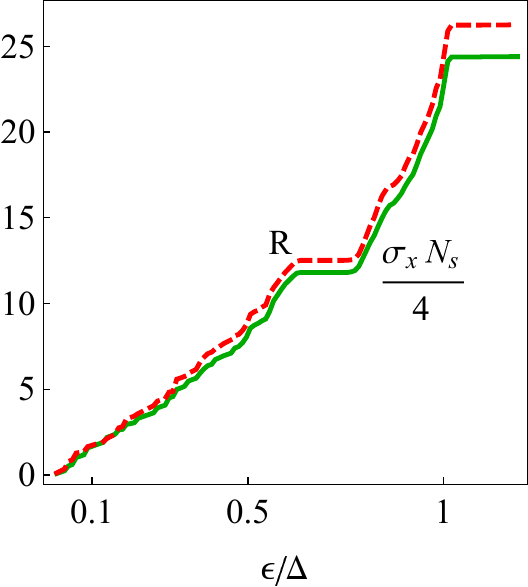} %[width = 1 \columnwidth, unit=1pt]
\put(55,105){$(a)$}
\end{overpic}
 \hspace{2 mm}
 \begin{overpic}[scale=0.96]{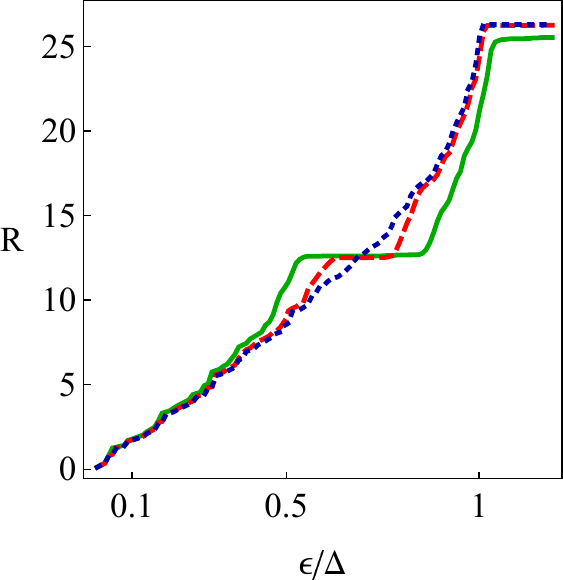} %[width = 1 \columnwidth, unit=1pt]
\put(55,105){$(b)$}
\end{overpic}

\vspace {8 mm}
 \begin{overpic}{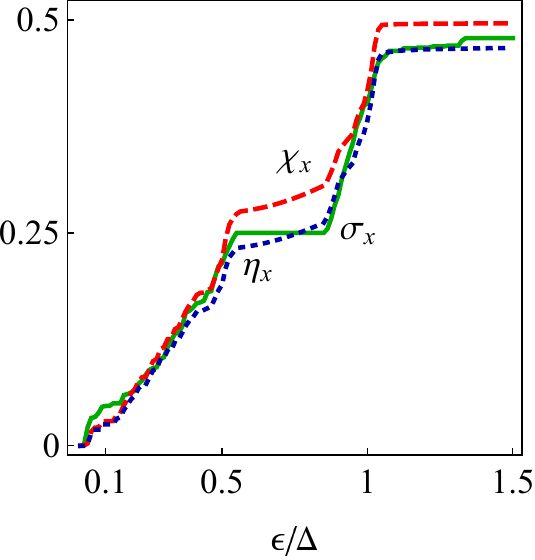} %[width = 1 \columnwidth, unit=1pt]
\put(55,105){$(c)$}
\end{overpic}
 \hspace{2 mm}
 \begin{overpic}{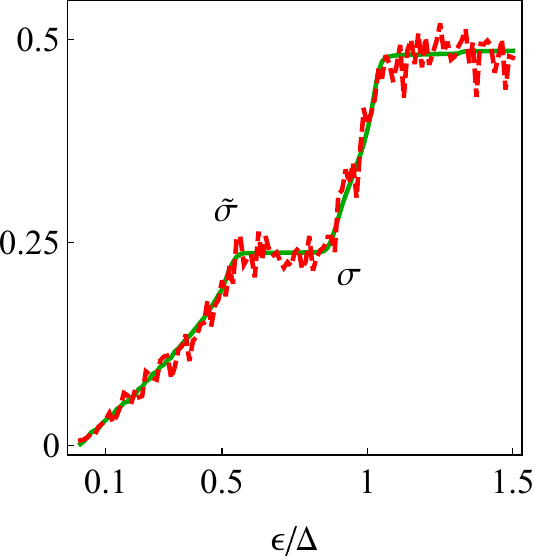} %[width = 1 \columnwidth, unit=1pt]
\put(55,105){$(d)$}
\end{overpic}
\end{center}
\caption{(a) The quantity $R$ in Eq.\ (\ref{radius}) (dashed red) and $\sigma_x N_s/4$ (solid green), in a system with $N_s=200$ and $J=0.05 \Delta$ and the impurity placed in the centre of the lattice. 
(b) $R$ versus $\epsilon/\Delta$, for $J/\Delta=0.01,0.05,0.1$ (dotted blue, dashed red and solid green respectively).
(c) Comparison among the quantities $\sigma_x$ 
in Eq.\ (\ref{measure_single}), solid green, 
$\chi_x$ in Eq.\ (\ref{chi_position}), dashed red, and $\eta_x$ in Eq.\ (\ref{eta_position}), dotted blue.
The impurity is at the centre of the lattice, $N_s=200$, $J=0.1 \Delta$. 
(d) probability $\tilde \sigma (\epsilon)$, in Eq.\ (\ref{sigma_rand}),
to generate an orthogonality event when every realization of the phase is associated with a random position of the impurity 
(dashed red); for comparison, we display in solid green the case in which the position is averaged over the whole lattice at every random realization of the phase.  In all the panels $\beta$ is the golden ratio. }
\label{singlestat}
\end{figure*}

\subsection {Noise sources and density measurements}
We shall now corroborate our analysis by looking at other interesting quantities.
Let us first observe that the function $\sigma$ defined in Eq.\ (\ref {measure}) is obtained by averaging over all possible positions of the impurity \emph{and} over the random phase of the quasi-periodic potential. 
Let us consider the following quantity
\begin{equation}
\sigma_x (\epsilon) = \left\langle \theta(\delta - F(x,\epsilon)) \right\rangle_\textrm{noise},
\label{measure_single}
\end{equation}
where the position $x$ of the impurity is fixed and the average over the lattice sites is not performed.
In the previous section we saw that an orthogonality event is generated whenever the energy of a site, as a consequence 
of the perturbation, becomes larger than the Fermi energy. Being the system highly localized, this orthogonality event is therefore associated with a particle occupying, in the new ground state, the site relative to the Fermi energy of the impurity-free system. Such a site is spatially separated from the site left unoccupied in the new configuration. 
In other words, the orthogonality event is associated with a rearrangement of the occupied sites. Motivated by this reasoning, let us define
\begin{equation}
R_x (\epsilon) = \left\langle  \sum_{j=1}^{N_s} \left |(j-x)\left [n(j) -  \tilde n(j,\epsilon,x)\right ] \right| \right\rangle_\textrm{noise},
\label {radius}
\end{equation}
in which $n(j)= \langle \Psi_0(\epsilon=0)| a^\dagger_j a_j |\Psi_0( \epsilon=0) \rangle$ is the ground state occupation of the $j$-th site in the absence of the impurity, and $\tilde n(j,\epsilon,x)= \langle \Psi_0(x,\epsilon)| a^\dagger_j a_j |\Psi_0(x, \epsilon) \rangle$ 
 is the ground state occupation of the $j$-th site  in the presence of an impurity at site $x$ with interaction strength $\epsilon$. The latter can 
 be seen as the adiabatic response to the quench. The quantity $R$ can be roughly interpreted as the average distance at which a particle is adiabatically moved as a consequence of the perturbation. As explained above, the particle will move to the site corresponding to the Fermi energy of the system without impurity. This site, for random realisations of the phase, can correspond to any lattice site with uniform probability. Therefore, for an impurity placed at the centre of the lattice, the site corresponding to the Fermi energy
 will be at an average distance  $ \sim N_s/4$. By considering the full statistics of the adiabatic transport, as a function of the perturbation potential, we can write  $R (\epsilon) \simeq \sigma_x (\epsilon) N_s/4$, where we are roughly assuming that the probability of an orthogonality event is equivalent to the probability of adiabatically transfering a charge (notice that we have taken $x= N_s/2$ and dropped the label $x$). Figure \ref {singlestat}(a) strongly corroborates this assumption; notice also how the quantity 
 $\sigma_x(\epsilon)N_s/4$ saturates to a value $\simeq n N_s/4$ for large $\epsilon$. It is then not surprising to see that, as displayed in Fig.\ \ref {singlestat}(b), by varying the ratio $J/\Delta$, the plateau of the function $R$ displays the same qualitative features of the $\sigma$ function. This is in line with the findings of Ref.\ \cite{khemani2015nonlocal}, where it is shown that the radius of disturbance does not scale like the localisation length, which in turn is determined by the ratio $J/\Delta$, but rather grows linearly with $N_s$. 

In order to further analyse the changes induced by the impurity in the density profile, let us define 
\begin{equation}
 \chi_x (\epsilon) = \left\langle  \frac{1}{2} \sum_{j=1}^{N_s}  |n(j) - \tilde n(j,\epsilon,x)| \right\rangle_\textrm{noise},
\label {chi_position}
\end{equation}
and similarly
\begin{equation}
 \eta_x (\epsilon) = \left\langle \frac{1}{2}  \sum_{j=1}^{N_s} [n(j) - \tilde n(j,\epsilon,x)]^2 \right\rangle_\textrm{noise}.
\label {eta_position}
\end{equation}
These two quantities enable us to further link the statistics of orthogonality events to the statistics of adiabatic charge transfers induced by the local quench. Also, they should be easy to access experimentally,  as confirmed 
by recent experiment in cold atomic gas where a single site resolution has been successfully achieved \cite{haller2015}.
These two quantities mimic to some extent the behaviour of the probability $\sigma$ in Eq.~\eqref{measure_single}. Indeed, the quantity $\chi$ in Eq.~\eqref {chi_position}, for a single realisation of the random phase, is bounded to take values in the interval $[0,1]$. The limiting cases are easily understood: $\chi=0$ when the two density profiles coincide, and $\chi=1$ when the density differs only for two spatially separated states. In this scenario, $\chi$ in Eq.~\eqref {chi_position} acts as a witness of the density rearrangement, and in turn as a signature for an orthogonality event. Furthermore, for strong values of the perturbation, it saturates to the filling factor. An analogous reasoning can be done for the quantity introduced in Eq.~\eqref {eta_position}.

As shown in Fig.\ \ref {singlestat}{(c)}, the $\sigma_x$ function displays a trend similar to $\sigma$ averaged over the lattice sites. Furthermore, both $\chi_x$ and $\eta_x$ also show an anomalous behaviour in the plateau region, where they
increase with a slower rate in $\epsilon$. The intuitive explanation of this behaviour is that when $\epsilon$ is increased in this interval, the number orthogonality events, or charge transfer events, does not change  but the wave function of the perturbed site is nonetheless modified.
In order to test the robustness of this phenomenon with respect to impurity position, we assume that 
at each realisation of the quasi-periodic potential the impurity is plunged at a completely random site. We define
\begin{equation}
\tilde \sigma (\epsilon) = \left\langle \theta(\delta - F( x_\phi ,\epsilon)) \right\rangle_\textrm{noise},
\label{sigma_rand}
\end{equation}
in which $ x_\phi$ represents the position of the impurity associated to each realization of the quasi-periodic potential, characterized by a random phase $\phi$. Despite this further source of noise we see that also $ \tilde \sigma (\epsilon)$ features a plateau, as shown in Fig.\ \ref {singlestat}{(d)}. 

\begin{figure*}[!t]
 \begin{center}
\vspace {5 mm}
 \begin{overpic}{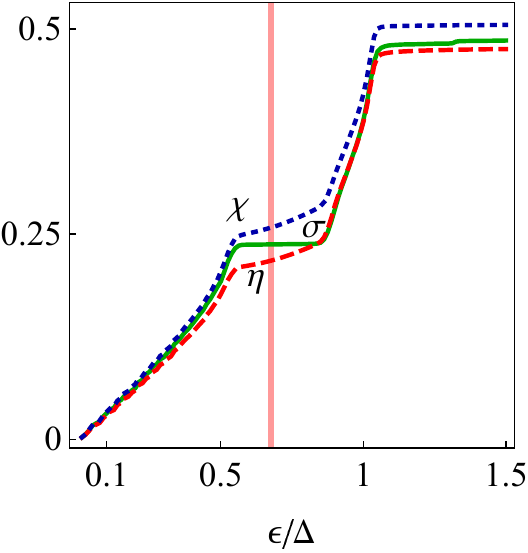} %[width = 1 \columnwidth, unit=1pt]
\put(55,105){$(a)$}
\end{overpic}
 \hspace{1 mm}
 \begin{overpic}{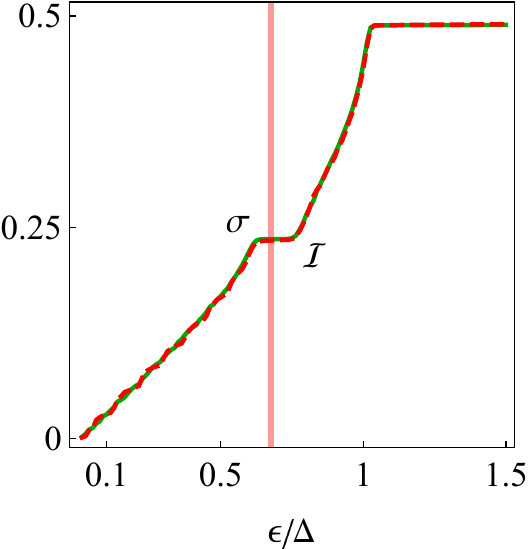} %[width = 1 \columnwidth, unit=1pt]
\put(55,105){$(b)$}
\end{overpic}

\vspace {8 mm}
 \begin{overpic}{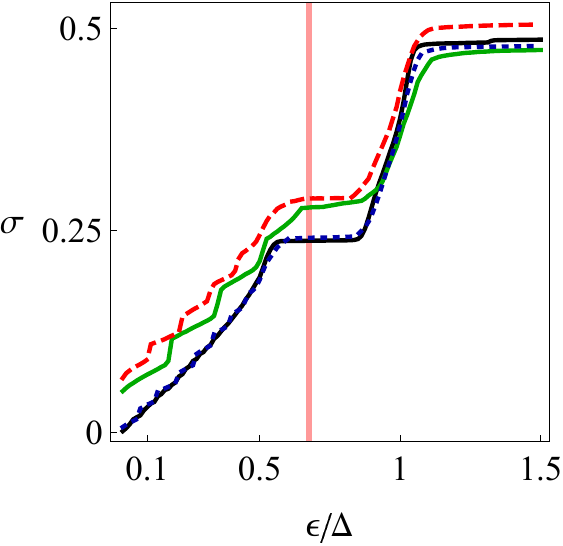} %[width = 1 \columnwidth, unit=1pt]
\put(55,100){$(c)$}
\end{overpic}
 \begin{overpic}{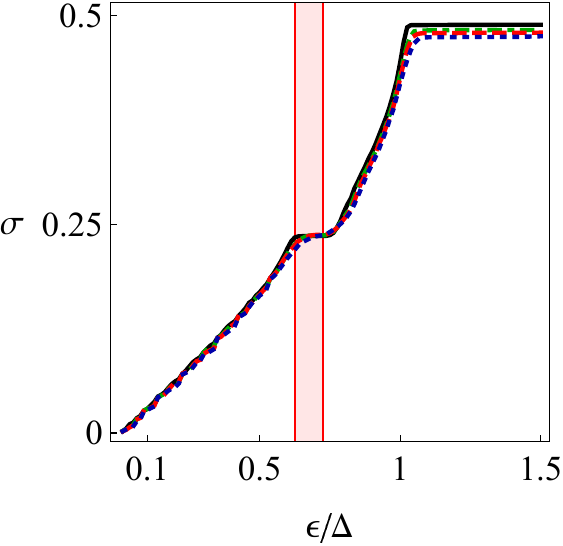} %[width = 1 \columnwidth, unit=1pt]
\put(55,100){$(d)$}
\end{overpic}

\end{center}
\caption{
(a) The quantities $\chi$ in Eq.\ (\ref{chichi}) (dotted blu) and $\eta$ in Eq.\ (\ref{etaeta}) (dashed red) are compared with $\sigma$ (solid green), in a system with $N_s=200$ and $J=0.1 \Delta$, $\beta$ being the golden ratio. (b)  $\mathcal I$ in Eq.\ (\ref{imbalance}) (dashed red) compared with $\sigma$  (solid green), in a system with $N_s=200$ and $J=0.05 \Delta$, $\beta$ being the golden ratio. (c) $\sigma$ is displayed for different approximation of the golden ratio with $\beta = \frac {55}{34},\frac {89}{55},\frac {144}{89}$, in green, red and blue respectively, with $N_S=100$ and $J=0.1 \Delta$. In solid black the $\sigma$ function with $\beta = \frac {1+ \sqrt 5}{2}$. (d) $\sigma$ is displayed for different lattice size, $N_S=200,120,100,80$, in black, green, red and blue respectively with $J=0.05 \Delta$ and $\beta$ being the golden ratio. In (a), (b) and (c) the vertical red line is at  $\epsilon/\Delta=|\sin(2 \pi\beta)|$. In (d) the two vertical lines are at  $\epsilon/\Delta=|\sin(2 \pi\beta)|-J/\Delta$ and  $\epsilon/\Delta=|\sin(2 \pi\beta)|+J/\Delta$. }
\label{finitesizeplot}
\end{figure*}

As a further example, providing a more complete analysis of quench-induced changes in the density, we introduce the following quantities
\begin{equation}
 \chi = \frac {1} { N_s} \sum_{x=1}^{N_s}   \chi_x (\epsilon),
 \label{chichi}
\end{equation}
and
\begin{equation}
 \eta = \frac {1} { N_s} \sum_{x=1}^{N_s}   \eta_x (\epsilon),
 \label{etaeta}
\end{equation}
that correspond to averaging \eqref {chi_position} and \eqref {eta_position} over all possible impurity positions in the lattice. This further average smoothens the quantities define above, as displayed in Fig.\ \ref {finitesizeplot}(a), without altering the behaviour observed in the single site analysis. 

Finally, to show the versatility of the figures of merit based on measurements of the density profile, we provide another possible quantifier, that is built by using the density imbalance between odd and even sites. Recent studies proved that the density imbalance can be efficiently monitored and can be employed to study relaxation properties in interacting Aubry-Andr\'e and many-body localisation phenomena \cite {schreiber2015observation,luschen2016}. Therefore, we look at
\begin{equation}
\mathcal I(\epsilon) = \frac{2}{N_s}\sum_{x=1}^{N_s}\left\langle \tilde n_{odd}(\epsilon, x)- n_{odd}(\epsilon=0) \right\rangle_\textrm{noise},
 \label{imbalance}
\end{equation}
where, $\tilde n_{odd}(\epsilon, x)$ is the occupation of all odd sites when the impurity is at site $x$ with interaction strength $\epsilon$. We monitor then the fluctuation over the occupation of the odd sites due to the presence of the impurity, by averaging as usual over the random realisation of the quasi-periodic potential and the position of the pertubation.
In Fig.\ \ref {finitesizeplot}(b) we see that this quantifier is able to capture the plateau of 
the $\sigma$ function. Such a good overlap can be explained by considering that an orthogonality 
event is associated to a charge transfer in the lattice that alters the number of odd (and even) occupied sites.

\subsection {Interferometric protocol}
Recent developments in the physics of impurity-based quantum probing techniques suggest an alternative, yet viable, scheme to measure the fidelity in Eq.\ \eqref {gsfidelity} directly (see \cite{PhysRevX.2.041020}). The key idea is to associate an additional energy level to the impurity acting as local perturbation and to design an interaction for which just one degree of freedom of the impurity couples to the fermionic bath, e.g. $\epsilon (t) \ket e \bra e \otimes a_x^\dagger a_x$ (for the sake of simplicity the impurity is modelled as a two level system with levels $\ket g$ and $\ket e$). After intialising the impurity state in an equal superposition $\frac {1}{2} (\ket g +\ket e)$, we assume  to adiabatically couple it to the gas. The evolution generated by the Hamiltonian $H$ will not lead to population transfer,  but it will cause dephasing of the impurity coherences. The asymptotic value of the off-diagonal elements of the impurity density matrix leads to the following fidelity 
\begin {equation}
|\rho_{eg} (+\infty) |=  |\langle \Psi_0(\epsilon=0)| U(+\infty) |\Psi_0(\epsilon=0) \rangle |=|\langle \Psi_0(\epsilon=0) |\Psi_0(x,\epsilon) \rangle|.
\end{equation}
By analysing then the statistics of the probe coherences after the adiabatic coupling, it should be possible to reconstruct the $\sigma (\epsilon)$ function
in Eq.\ (\ref{measure}). This interferometric approach has been experimentally realised to study the dynamics of impurities coupled to a Fermi sea \cite{cetina2016ultrafast}, and it has been employed extensively in the attempt to design quantum probing protocols for cold trapped atoms, {\it e.g.} to measure a gas temperature  \cite {johnson2016}, quantum correlations in bosonic systems  \cite {elliott2016,streif2016}, and to probe the orthogonality catastrophe in trapped fermionic environments \cite {schmidt2018universal,sindona2013}.

\subsection {Lattice size and incommensurability}
We finally discuss the role of the lattice size in the emergence of the plateau structure. 
 The size of the system will play an important role when realising a quasi-periodic potential, as it determines when a good approximation of the irrationality of the incommensurate frequency is achieved. If $F_n$ is the $n$-th element of the Fibonacci sequence, the ratio $\frac {F_{n+1}}{F_n}$ converges to the golden ratio in the $n \rightarrow \infty$ limit. Therefore, for a finite size lattice, a good approximation of the quasi-periodic potential is achieved whenever $\beta = \frac {F_{n+1}}{F_n}$, with $F_{n+1} \gtrsim  N_s$. Figure \ref{finitesizeplot}(c) displays the sigma function for different approximations of the golden ratio.

\section{Conclusion}
\label{conclusions}

We have explored the statistics of orthogonality catastrophe events resulting from adiabatically perturbing a system of non-interacting and strongly localised fermions in a disordered lattice. This has led to new and unexpected features directly linked to the very nature of the quasi-periodic potential. In particular, we have shown that the gapped, fractal spectrum of the Aubry-Andr\'e model and the energy-space resonances affect the statistical orthogonality catastrophe quite drastically, resulting in a plateau structure. We have also provided a connection with experimentally accessible quantities, based on either measurement of the lattice density profile, or atom impurities serving as controllable quantum probes, suggesting an experimental verification to be well within reach with current available technologies.

We also stress that our analysis and numerical simulations pertain to lattice sizes of  O$(10^2 \div 10^3)$, figure \ref{finitesizeplot}(d). This is the realm where most experiments can be performed. 
Increasing the lattice size will eventually reveal the scaling of the typical fidelity, as in Eq.~\eqref{stoc}. The study of the features and fine details of this transitions are left for a future investigation.

\section*{Acknowledgements}
F. C., M. B., E.-M. L. and S. M. acknowledge financial support from the Horizon 2020 EU collaborative project QuProCS (Grant Agreement 641277), the Academy of Finland Centre of Excellence program (Project no. 312058) and the Academy of Finland (Project no. 287750).
S. P. is partly supported by INFN through the project ``QUANTUM".
A. S. is partly supported by a Google Faculty Award. The computer resources of the Finnish IT Center for Science (CSC) and the FGCI project (Finland) are acknowledged.

%\bibliographystyle{apsrev4-1}
%\bibliography{stocbibl}
\section*{References}
\bibliography{MBLbib}

\end{document}